\documentstyle[11pt,newpasp,twoside,epsf]{article}
\def\edcomment#1{\iffalse\marginpar{\raggedright\sl#1\/}\else\relax\fi}
\reversemarginpar

\begin{document}
\title{A white dwarf with a low magnetic field in the IP RXJ0028.8+5917/V709Cas} 

\author{J.M. Bonnet-Bidaud(1), 
M. Mouchet (2)
D. de Martino (3) 
G. Matt (4)
\& C. Motch (5)} 

\affil{(1)Service d'Astrophysique, DSM/DAPNIA/SAp, CE Saclay, 
F-91191 Gif sur Yvette Cedex, France,
(2) DAEC et UMR 8631 du CNRS, Observatoire de Paris Section de Meudon, F-92195 Meudon Cedex, France,
(3) Osservatorio Astronomico di Capodimonte, I-80131 Napoli, Italy,
(4) Dipartimento di Fisica, Universita Roma Tre, I-00146 Roma, Italy,
(5) CNRS,  Observatoire de Strasbourg, F-67000 Strasbourg, France }

\begin{abstract}
We report the first detailed spectroscopic observations of the recently 
identified intermediate polar RXJ0028.8+5917/V709 Cas. We discovered 
that the system shows significant EW $\sim$(2-4)\AA\, broad absorptions affecting 
the Balmer lines from H$\delta$ to H$\beta$. 
These broad absorptions are interpreted as the contribution of an underlying 
DA log\,g=8 white dwarf at a temperature of $\sim$23 000 K, contributing  $\sim$17 \% 
(at 4500 \AA) to the overall flux. 
This is the first direct detection of a white dwarf in an intermediate polar system.
The absence of significant Zeeman splitting indicates a magnetic field lower 
than 10 MG, confirming that, at least in some cases, intermediate polars 
have weaker fields than polars.
\end{abstract}

\section{Introduction}
The X-ray source RXJ0028.8+5917 was discovered by ROSAT as an Intermediate 
Polar with a 312 sec X-ray pulsation and was identified with a m$_v$=14 variable 
star V709 Cas (Haberl \& Motch 1995, Motch et al. 1996). It was also extensively 
observed with  the BeppoSAX satellite  (de Martino et al., this conference).
Observations of RXJ0028.8+5917/V709 Cas were performed at the 1.93m telescope of the Haute-Provence 
Observatory (OHP, France) during 4 nights on Aug. 1998.
43 spectra with 7 \AA\, resolution and 15 or 20 min exposures were obtained in 
the range (3600-7200\AA).
The optical spectrum is typical of IPs with strong H and He emission lines.
Radial velocities on the 4 nights were used to determine the best orbital 
ephemeris  of the blue-to-red crossing time as :
To = HJD 2451048.0575(2) + E*0.2225(2) d
(or P = (5.341 $\pm$ 0.005) h). 
This removes the previous uncertainty between different aliases. 

\section{A two-component spectrum}
The mean optical spectrum of V709 Cas also shows 
clear broad absorptions features in the Balmer lines, H$\delta$,  
H$\gamma$ and  H$\beta$, while such a feature is absent around HeII
(see Bonnet-Bidaud et al. 2001, for the figures).
Such absorption lines were not yet seen in any other IPs.
Lines in absorption are sometimes seen among classical CVs, 
mainly in nova-like systems in high states and during dwarf nova 
eruptions, which suggests that they are formed in an optically thick 
disk with a high mass transfer (see for instance La Dous 1994, 
Hessman 1986). In the case of V709 Cas, the stability of the optical flux 
excludes a dwarf nova event and the absence of He absorptions does not 
favour a disk origin. 	
The absorptions are interpreted as coming from the white dwarf atmopshere. 
The line FWHMs have been compared to a grid of white dwarf models 
(Koester 2000) and the line EWs have been used to determine the white 
dwarf contribution to the overall flux.
The range of values of the measured FWHM absorptions (51-65 \AA) is found consistent 
with a log\,g=8.0 white dwarf at a temperature of (18 000 - 30 000 K) 
with a best value at T = 23 000 K. 
The white dwarf is found to contribute $\sim$17\% at 4500\AA\, and only $\sim$6\% at 
6500\AA. The comparison of the measured to theoretical flux values yields a
Rwd(10$^{9}$cm/D(pc)) value of (0.35-0.42), corresponding to a distance D=(210-250pc), 
for the (18 000-30 000K) temperature range (see Bonnet-Bidaud et al. 2001 for details).

\section{Discussion}
What is the magnetic field value in V709 Cas ? Absorption lines are commonly seen in polars 
during low states where they are split by the Zeeman effect of a 
(10-30MG) strong magnetic field.
The absence of Zeeman splitting in V709 Cas indicates a low magnetic field.
Comparaison has been made with synthetic profiles for B in the range 
(3-30)MG  (see Bonnet-Bidaud et al. 2000). The absence of the H$\beta$ 
components at 4832\AA\, and 4896\AA\,  imposes B $\leq$ 3 MG.
This is the first direct evidence of a low magnetic field, at least in some IPs.\\
Why is the white dwarf visible in V709 Cas ?
V709 Cas is the only IP where the WD contribution is significant.
The white dwarf characteristics are not atypical among CVs.
A high (WD/overall) flux ratio may be reached if the contribution 
from other regions is significantly lower than in other IPs.  
This could be the case if the accretion disk is seen at high inclination.
Additional observations are clearly needed to see if the visibility of the 
white dwarf may also depend on the overall accretion rate and luminosity of the source. 

\section{References}
Bonnet-Bidaud, J.M., Mouchet, M., Shakhovskoy, N. M., Somova, T. A., 
Somov, N. N. et al. 2000, A\&A 354, p.1003\\ 
Bonnet-Bidaud, J.M., Mouchet, M., de Martino D., Matt, G., Motch, C. 2001, A\&A 374, 1003\\ 
Haberl, F., Motch, C. 1995, A\&A 297, 37\\
Hessman, F.V., Koester, D., Schoembs, R., Parwigh, H. 1989, A\&A 213, 167\\ 
Koester, D. 2000, private communication\\
La Dous, C. 1994, Sp Sci Rev. 67, p. 1\\ 
Motch, C., Haberl, F., Guillout, P., Pakull, M., Reinsch, K. 1996, A\&A 307, 459\\ 

\end{document}